\documentclass{article}
\usepackage{spconf,amsmath,graphicx}
\usepackage{booktabs}
\usepackage{xcolor}

\title{EXPLORING TRANSFORMER'S POTENTIAL ON AUTOMATIC PIANO TRANSCRIPTION}

\name{
    Longshen Ou\textsuperscript{1} \qquad
    Ziyi Guo\textsuperscript{1} \qquad 
    Emmanouil Benetos\textsuperscript{2} \qquad 
    Jiqing Han\textsuperscript{3} \qquad 
    Ye Wang\textsuperscript{1}
}
\address{Author Affiliation(s)}
\address{
    \textsuperscript{1}School of Computing, National University of Singapore, Singapore\\
    \textsuperscript{2}Centre for Digital Music, Queen Mary University of London, UK\\
    \textsuperscript{3}School of Computer Science and Technology, Harbin Institute of Technology, China
}
\begin{document}

\maketitle

\ninept
\begin{abstract}

Most recent research about automatic music transcription (AMT) uses convolutional neural networks and recurrent neural networks to model the mapping from music signals to symbolic notation. Based on a high-resolution piano transcription system, we explore the possibility of incorporating another powerful sequence transformation tool---the Transformer---to deal with the AMT problem. We argue that the properties of the Transformer make it more suitable for certain AMT subtasks. We confirm the Transformer’s superiority on the velocity detection task by experiments on the MAESTRO dataset and a cross-dataset evaluation on the MAPS dataset. We observe a performance improvement on both frame-level and note-level metrics after introducing the Transformer network.

\end{abstract}

\section{Introduction}
\label{sec:intro}

Automatic music transcription (AMT) aims to convert music signals into music notation. It is of great importance to solve the AMT problem because the transcription results can be helpful in many higher-level tasks, like structure segmentation, music similarity assessment, and so on \cite{Benetos2019}. However, it is not easy to provide a generic solution to AMT, since a music piece usually contains multiple different sound sources and lots of simultaneous notes. It is usually difficult to separate these polyphonic sounds.

Piano transcription is one subproblem of AMT. Because automated annotation tools (e.g., Yamaha Disklavier) can be used to help us capture the annotation of music data, we have relatively rich datasets for the piano transcription problem, compared to other instruments. This advantage makes it possible to use powerful supervised learning approaches, of which neural networks (NNs) are a representative group of methods. 

Recently, the Transformer networks \cite{Vaswani2017} have gathered researchers’ attention from different fields. The success of Music Transformer \cite{Huang2019} shows the superiority of using Transformers to model symbolic music. Given the great potential of the Transformer on sequence modeling, we would like to explore its ability to solve AMT tasks, where both input and output sequences have finer granularity than \cite{Huang2019}. Furthermore, because of its ability to model relationships between all time steps in a sequence, it is especially good at modeling long-term dependencies. We expect that this property could bring positive impacts when dealing with AMT tasks.

To test the modeling ability of the Transformer for AMT, in this paper, we attempt to use Transformers to solve various subtasks of piano transcription, including multi-pitch detection, onset and offset detection, and velocity estimation, to explore the possible improvement, and find that the Transformer gives a relatively significant improvement on velocity detection task. Then, based on the system in \cite{Kong2020amt}, we try to incorporate the Transformer into the velocity branch to enhance the performance of velocity detection. We train and evaluate our system using the MAESTRO dataset \cite{Hawthorne2019} and perform a cross-dataset evaluation using the MAPS dataset \cite{Emiya2010}. We notice that the improvement of the velocity branch has positive effects on the overall performance of the transcription system, by providing more accurate information to the downstream modules of the velocity branch, i.e., onset and frame branches. The proposed system achieves competitive results compared to previous state-of-the-art systems.

\section{related work}

The first trial of the NN-based AMT method starts with \cite{Marolt2004}, in which the feasibilities of several basic network structures were tested, including recurrent neural networks (RNNs). B{\"{o}}ck et al. \cite{Bock2012} tried to apply Long Short-Term Memory (LSTM) NNs to solve AMT tasks. Sigtia et al. \cite{Sigtia2016} used convolutional neural networks (CNNs) as the acoustic model, and integrated an RNN-based music language model to improve the performance further. Hawthorne et al. \cite{Hawthorne2018} tried to do note-level transcription by designing networks for onset detection and using the onset information to help the learning of the multi-pitch estimation network. Kelz et al. \cite{Kelz2019} tried to model the time-variant note properties by considering different note stages. Kim et al. \cite{Kim2019} introduce an adversarial training scheme for NN-based methods to more accurately express inter-label dependencies. Kong et al. \cite{Kong2020amt} designed a network that can provide more refined transcription results containing onset, offset, note pitch, and key velocity (speed of pressing a key). More recently, Hawthorne et al. \cite{Hawthorne2021} shows that a generic Transformer without domain-specific adaptation can be used to generate note-level transcription results directly with competitive performance. 

Beyond pitch and timing information, dynamics (referred to as `intensities' or `velocities' interchangeably) are another important factor of music. Szeto et al. \cite{Szeto2005} proposed to search velocity value by employing a single-note database to artificially generate mixtures of notes. A parametric spectrogram model to estimate note intensities was proposed in \cite{Ewert2011}. Van Herwaarden et al. \cite{VanHerwaarden2014} tried to utilize Restricted Boltzmann Machines to deal with this task. Methods in \cite{Jeong2017} and \cite{Jeong2018} are based on non-negative matrix factorization. However, most of the previous research was done in a score-informed manner and is not conducted in the AMT context. The first trial of estimating note dynamics alongside the pitch and timing information is \cite{Hawthorne2018}. Models in \cite{Kong2020amt} and \cite{Hawthorne2021} significantly improve the NN-based velocity estimation performance by more effective network structure.

The development of deep learning has brought more powerful tools for NN-based methods in different fields. The Transformer has become a revolutionary architecture in recent years. One of its important internal components is the self-attention mechanism. It allows modeling of dependencies without regard to their distance in the input or output sequences \cite{Vaswani2017}. By using the attention mechanism entirely and eschewing recurrent structures, it can capture global dependencies between input and output sequences, and meanwhile, allow more parallel computing than RNNs, hence improving the training efficiency. 

\section{method}
\label{sec:method}
\subsection{System architecture}
\label{subsec:system}

\begin{figure}[t]

\begin{minipage}[b]{1.0\linewidth}
  \centering
  \centerline{\includegraphics[width=5.3cm]{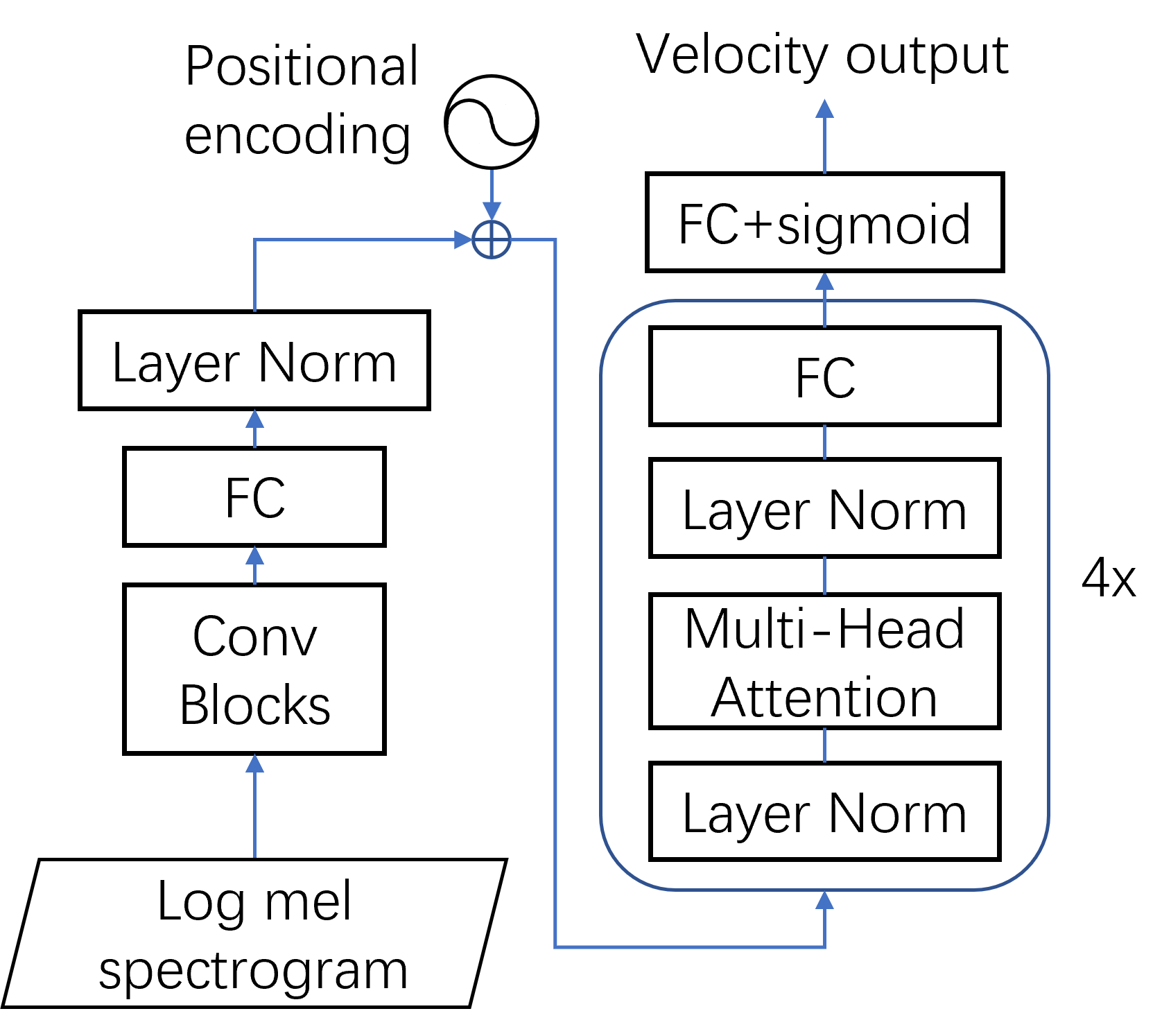}}
\end{minipage}

\caption{Architecture of the CNN-Transformer.}
\label{fig:trans}

\end{figure}

The architecture of the proposed piano transcription system is based on the high-resolution piano transcription system in \cite{Kong2020amt} (hereinafter called the baseline system). Inherent from the baseline system, we divide the entire task into four subtasks, i.e., onset detection, offset detection, multi-pitch estimation (frame classification), and velocity estimation. We use the same data preprocessing and postprocessing method as the baseline system. 

\begin{figure}[t]

\begin{minipage}[b]{1.0\linewidth}
  \centering
  \centerline{\includegraphics[width=5.7cm]{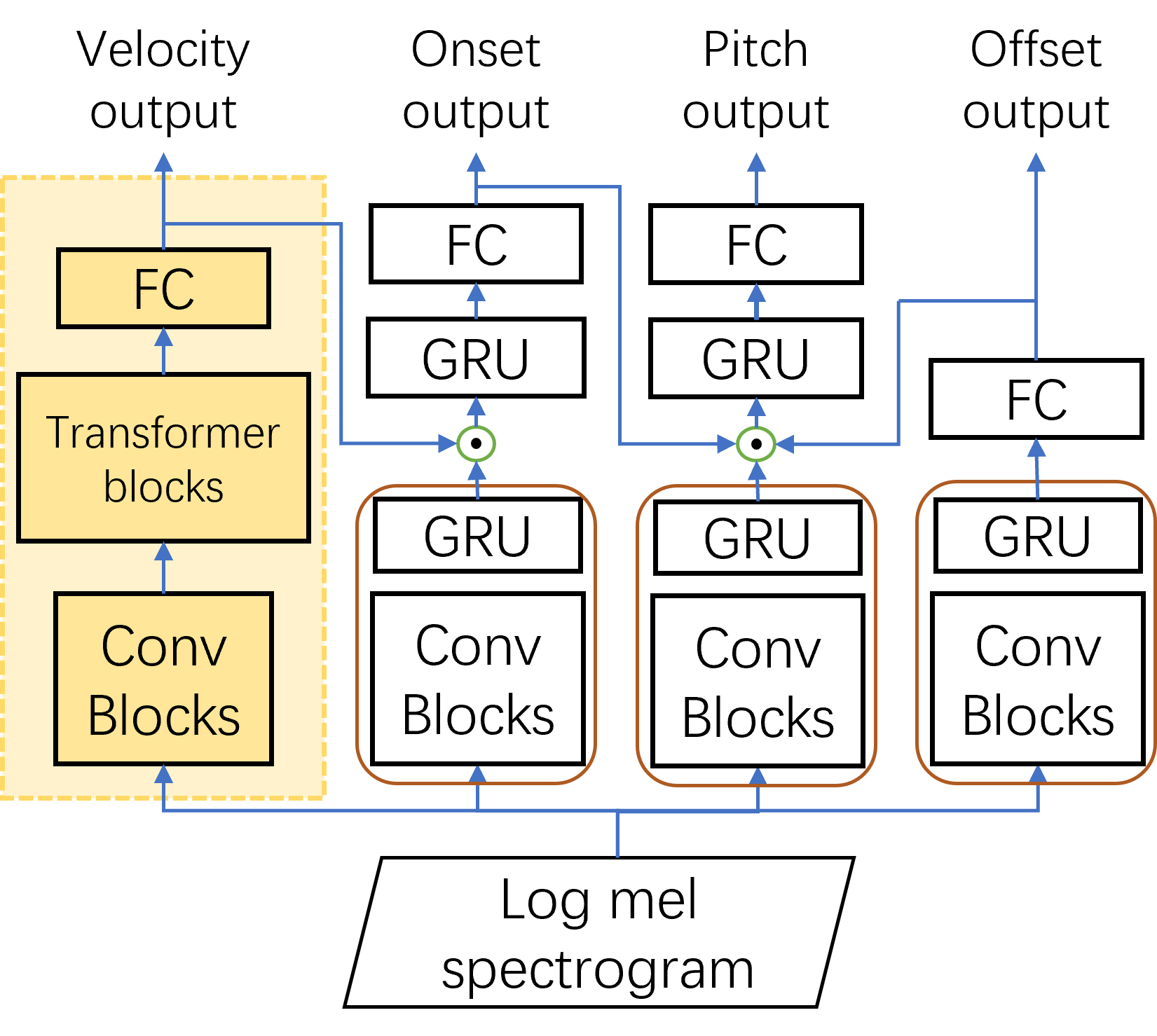}}
\end{minipage}

\caption{Architecture of the proposed system. The highlighted part is the proposed CNN-Transformer; the other parts are inherited from the baseline system. The $ \odot $ refers to concatenation operation.}
\label{fig:system}

\end{figure}

Based on the framework of the baseline system, we first attempt to substitute the bi-directional gated recurrent unit networks (GRU) in the original system with the Transformers to see if there is any performance improvement. The resulting CNN-Transformer architecture is shown in Fig. \ref{fig:trans}. We try this architecture on each subtask of the original system, evaluate its performance, and compare it with the baseline system. Among these experiments, we observe a considerable performance improvement on the velocity estimation tasks. Hence for the final transcription system, we use the CNN-Transformer structure for the velocity estimation branch and keep the CNN-GRU combination for other branches. The architecture of the resulting system is illustrated in Fig. \ref{fig:system}. The detailed structure of the CNN-Transformer for velocity estimation is included in Section \ref{sec:transformer}.

\subsection{Incorporating the Transformer}
\label{sec:transformer}

Following the baseline system, we solve the AMT problem using a two-step approach: we first convert a sequence of discretized music signals into a sequence of frame-level symbolic notations using the designed neural networks, and then perform postprocessing for these frame-level results by a note search approach to get the final note-level results. The sequence-to-sequence model can be easily incorporated into the first step in this scenario. 

Fig. \ref{fig:trans} shows the structure of the network that aims to solve one subtask of AMT. The overall design principle is to let the Transformer concentrate on capturing long-range dependencies, by inserting convolutional layers at the beginning of the network that tries to learn short-term dependencies. Before input into the network, the original data is first resampled to 16 kHz, split into 10-second clips, and converted into log-mel spectrograms, with 229 mel bins, a Hanning window with size 2048, and a hop size of 10 ms. Then, we use a stacked CNN to extract features of the spectrogram. This process can be seen as learning local relationships with convolutional layers within a relatively small context. The detailed setup of the CNN is inherited from \cite{Kong2020amt}, which is shown in Tables \ref{tab:inside_conv_block} and \ref{tab:conv_blocks}. Then, the features outputted by CNN are reduced to the same dimension as the output of Transformer blocks by a position-wise fully connected layer, followed by a layer normalization operation, where the architecture of the CNN-Transformer diverges from the CNN-GRU. After that, we compute sinusoidal positional encoding as in \cite{Vaswani2017} and add it to the data. Subsequently, a series of stacked Transformer blocks is used to capture the long-range relationships for the sequence. Each block contains a multi-head attention layer, a fully connected layer, and two layer normalization operations. 

\begin{table}[b]
\centering
\resizebox{0.88\linewidth}{!}{%
\begin{tabular}{@{}ccccc@{}}
\toprule
\textbf{Layer} & \textbf{Output channel} & \textbf{Filter size} & \textbf{Stride} & \textbf{Padding} \\ \midrule
Conv2d & $m$ & 3x3 & 1 & 1 \\
BN 2d &  &  &  &  \\
Relu &  &  &  &  \\
Conv2d & $m$ & 3x3 & 1 & 1 \\
BN 2d &  &  &  &  \\
Relu &  &  &  &  \\
Avg pool 2d &  & 1 x 2 & 2 & 0 \\ \bottomrule
\end{tabular}%
}
\caption{Detailed structure of a convolution block with $m$ input channels.}
\label{tab:inside_conv_block}
\end{table}
\begin{table}[tb]
\centering
\resizebox{\linewidth}{!}{%
\begin{tabular}{@{}cccc@{}}
\toprule
\textbf{Layer} & \textbf{Input channel} & \textbf{Output channel} & \textbf{Output shape} \\ \midrule
Spectrogram input &  &  & 1 x 1001 x 229 (mel bins) \\
ConvBlock1 & 1 & 48 & 48 x 1001 x 114 \\
Dropout, p=0.2 &  &  &  \\
ConvBlock2 & 48 & 64 & 64 x 1001 x 57 \\
Dropout, p=0.2 &  &  &  \\
ConvBlock3 & 64 & 96 & 96 x 1001 x 28 \\
Dropout, p=0.2 &  &  &  \\
ConvBlock4 & 96 & 128 & 128 x 1001 x 14 \\
Dropout, p=0.2 &  &  &  \\ \bottomrule
\end{tabular}%
}
\caption{Detailed configuration of convolution blocks.}
\label{tab:conv_blocks}
\end{table}


Our CNN-Transformer model processes the input signal in a frame-by-frame manner. As in \cite{Hawthorne2018} and \cite{Kong2020amt}, classification or regression is directly performed on each frame of the spectrogram, which is a straightforward idea to recognize the note elements at every time step accurately. Since the time unit of the model's output is the same as that of input, and all the outputs are of fixed length, we did not utilize the original encoder-decoder structure of the Transformer.

\section{Evaluation}
\label{sec:eval}

\subsection{Dataset}

We train our model on the MAESTRO V3.0.0 dataset \cite{Hawthorne2019}. It contains recordings and corresponding MIDI files of the International Piano-e-Competition, which consists of over 200 hours of solo piano recordings. 
We use the train/validation/test split configuration proposed by the dataset provider to ensure that the same composition only appears in one subset.

To further show the superiority of our methods, we do a cross-dataset evaluation using the ``ENSTDkAm'' and the ``ENSTDkCl'' subsets of the MAPS database \cite{Emiya2010}. In this case, the system to be evaluated is still trained on the MAESTRO dataset.

\subsection{Evaluation measures}

We compute two groups of evaluation metrics to evaluate our model: metrics for subtasks and metrics for overall performance. Computation of these measures is implemented using {\fontfamily{pcr}\selectfont mir\_eval} \cite{Raffel2014}. 

To evaluate the performance on subtasks, we first compute frame-level metrics directly using the output of different branches. Following previous work \cite{Kong2020amt} and \cite{Hawthorne2018}, we use a frame size of 10 ms for evaluation. For the outputs of onset, offset and pitch branches, we compute the frame-wise F1 score as the measures. For the velocity branch, following \cite{Ewert2011} and \cite{Jeong2017}, we compute the mean and standard deviation of absolute error for frames on which there exist onsets. In addition to the frame-level metrics, we also compute windowed F1 scores for only timing information when evaluating onset and offset results. In this case, the correctness is determined in a small window. The window size (time tolerance) of both onset and offset is 50 ms. 

Note-level metrics are used to evaluate the overall performance of the system. We compute precision, recall, and F1 score for three types of note-level metrics, including note (onset and pitch), note with offset, and note with offset and velocity. When computing note-wise measures, we use 50 ms for onset and offset tolerance, and $10\%$ of velocity range for velocity tolerance, following the work of \cite{Hawthorne2018} and \cite{Kong2020amt}.

\begin{table}[!b]
\centering
\resizebox{\linewidth}{!}{%
\begin{tabular}{@{}cccccccc@{}}
\toprule
                & \textbf{Frame} & \multicolumn{2}{c}{\textbf{Onset}} & \multicolumn{2}{c}{\textbf{Offset}} & \multicolumn{2}{c}{\textbf{Velocity}} \\ \midrule
                & \textbf{F1}    & \textbf{F1-s}    & \textbf{F1-t}   & \textbf{F1-s}    & \textbf{F1-t}    & \textbf{MAE}      & \textbf{STD}      \\ \midrule
CNN-GRU         & \textbf{80.92} & 54.32            & \textbf{95.53}  & \textbf{27.98}   & \textbf{80.51}   & 4.2725            & 4.1925            \\
CNN-Transformer & 77.38          & \textbf{54.36}   & 95.18           & 22.66            & 73.87            & \textbf{4.0026}   & \textbf{4.0077}   \\ \midrule
Without PE      & 74.07          & 52.51            & 94.97           & 22.47            & 73.12            & 4.4241            & 4.3072            \\ \bottomrule
\end{tabular}
}
\caption{Performance of separately trained CNN-Transformers. F1-s refers to the strict frame-level F1 score. F1-t refers to windowed F1 score of timing information. The third row shows the results of CNN-Transformers without positional encoding. The lower velocity measures and higher other metrics indicate better performance.}
\label{tab:res-subtasks}
\end{table}
\begin{table}[hbt!]
\centering
\resizebox{\linewidth}{!}{%
\begin{tabular}{@{}cccccccc@{}}
\toprule
 & \textbf{Frame} & \multicolumn{2}{c}{\textbf{Onset}} & \multicolumn{2}{c}{\textbf{Offset}} & \multicolumn{2}{c}{\textbf{Velocity}} \\ \midrule
 & \textbf{F1} & \textbf{F1-s} & \textbf{F1-t} & \textbf{F1-s} & \textbf{F1-t} & \textbf{MAE} & \textbf{STD} \\ \midrule
Onset \& Frame {[}10{]} & 89.19 & - & - & - & - & - & - \\
Adversarial {[}12{]} & \textbf{91.40} & - & - & - & - & - & - \\
HPT {[}4{]} & 89.62 & - & - & - & - & - & - \\ \midrule
HPT reproduced & 89.63 & \textbf{57.79} & 97.23 & 31.79 & 87.68 & 3.2996 & 3.3822 \\
HPT-T & 90.09 & 57.44 & \textbf{97.31} & \textbf{32.28} & \textbf{88.44} & \textbf{2.9681} & \textbf{3.1583} \\ \bottomrule
\end{tabular}
}
\caption{Frame-level comparison.}
\label{tab:res-frame}
\end{table}

\begin{table*}[!htb]
\centering
\begin{tabular}{@{}cccccccccc@{}}
\toprule
 & \multicolumn{3}{c}{\textbf{Note-2}} & \multicolumn{3}{c}{\textbf{Note-3}} & \multicolumn{3}{c}{\textbf{Note-4}} \\ \midrule
 & \textbf{P} & \textbf{R} & \textbf{F1} & \textbf{P} & \textbf{R} & \textbf{F1} & \textbf{P} & \textbf{R} & \textbf{F1} \\ \midrule
Onset \& Frame {[}10{]} & 97.42\% & 92.37\% & 94.80\% & 81.84\% & 77.66\% & 79.67\% & 78.11\% & 74.13\% & 76.04\% \\
Adversarial {[}12{]} & 98.10\% & 93.20\% & 95.60\% & 83.50\% & 79.30\% & 81.30\% & 82.30\% & 78.20\% & 80.20\% \\
HPT {[}4{]} & 98.17\% & 95.35\% & 96.72\% & 83.68\% & 81.32\% & 82.47\% & 82.10\% & 79.80\% & 80.92\% \\
\multicolumn{1}{l}{Generic Transformer {[}13{]}} & \multicolumn{1}{l}{98.61\%} & \multicolumn{1}{l}{93.60\%} & \multicolumn{1}{l}{95.95\%} & \multicolumn{1}{l}{86.19\%} & \multicolumn{1}{l}{81.86\%} & \multicolumn{1}{l}{\textbf{83.46\%}} & \multicolumn{1}{l}{84.95\%} & \multicolumn{1}{l}{80.70\%} & \textbf{82.18\%} \\ \midrule
HPT reproduced & 98.00\% & 95.42\% & \underline{96.68\%} & 83.49\% & 81.33\% & 82.38\% & 81.99\% & 79.89\% & 80.91\% \\
HPT-T & 97.88\% & 95.72\% & \textbf{96.77\%} & 84.13\% & 82.31\% & \underline{83.20\%} & 82.85\% & 81.07\% & \underline{81.90\%} \\ \bottomrule
\end{tabular}
\caption{Note-level performance comparison. Note-2 refers to note (onset and pitch). Note-3 refers to note with offset. Note-4 refers to note with offset and velocity. The best F1 scores are bolded; the second-best are underlined. }
\label{tab:res-note}
\end{table*}


\begin{table*}[!tb]
\centering
\begin{tabular}{@{}ccccccccccc@{}}
\toprule
 & \textbf{Frame} & \multicolumn{3}{c}{\textbf{Note-2}} & \multicolumn{3}{c}{\textbf{Note-3}} & \multicolumn{3}{c}{\textbf{Note-4}} \\ \midrule
 & \textbf{F1} & \textbf{P} & \textbf{R} & \textbf{F1} & \textbf{P} & \textbf{R} & \textbf{F1} & \textbf{P} & \textbf{R} & \textbf{F1} \\ \midrule
HPT {[}4{]} & 82.77\% & 87.60\% & 84.23\% & \textbf{85.80\%} & 60.75\% & 58.48\% & 59.55\% & 45.80\% & 44.10\% & 44.90\% \\
HPT-T & \textbf{83.10\%} & 86.47\% & 85.13\% & 85.72\% & 60.97\% & 60.10\% & \textbf{60.48\%} & 47.40\% & 46.73\% & \textbf{47.02\%} \\ \bottomrule
\end{tabular}
\caption{Cross-dataset evaluation on MAPS dataset.}
\label{tab:res-cross}
\end{table*}

\subsection{Training}

The experiments consist of two parts. First, we used a subset of the MAESTRO dataset to train CNN-Transformer separately on four subtasks, i.e., onset and offset detection, multi-pitch estimation, and velocity estimation, to test the feasibility of using this structure for different tasks. We did not provide velocity information when detecting onsets using both CNN-GRU and CNN-Transformer; hence the architecture shown in \ref{fig:trans} is used for onset, offset, and velocity subtasks. When dealing with multi-pitch estimation subtask by CNN-Transformer, the onset and offset information is provided by CNN-GRU onset and offset branches. 

For the second part of the experiment, we take out the Transformer which performs well and use GRU for the rest of the branches to form the final transcription system, as shown in Fig \ref{fig:system}. We train the model using the ``train'' split of the MAESTRO dataset, and test the performance by MAESTRO's ``test'' split and two subsets of the MAPS dataset. During training, all branches in the system are jointly optimized.

Following the work in \cite{Hawthorne2018} and \cite{Kong2020amt}, we apply the sigmoid function to the output to scale them into the range of  $[0,1]$. We use binary cross-entropy loss for each subtask and their summation as the overall loss function. The model is optimized using the AdamW \cite{Loshchilov2019} algorithm. We use a batch size of 10 for all the experiments. For the Transformer network, we set the embedding dimension $d_{model}$ to 512, feed forward upward projection dimension $d_{ff}$ to 2048, and the head number to $4$. To avoid time-consuming learning rate adjustment and the overfitting issue, we use cyclical learning rate \cite{Smith2017} with ``triangular2'' strategy during the training process. We set the step size to 50k and train the model for 200k iterations.

\subsection{Results}

\subsubsection{Transformer on different subtasks}
\label{sssec:res_subtasks}

The experiments in this section use a subset of the MAESTRO dataset containing one-tenth of the original data. Table \ref{tab:res-subtasks} shows the results of applying the Transformer on different subtasks separately. On the onset task, the CNN-Transformer has a similar performance to CNN-GRU, with 0.04\% higher strict F1 and 0.35\% lower F1 with tolerance. On the velocity task, the Transformer has about 6.3\% lower mean absolute error and 4.4\% lower standard deviation of the absolute error than the baseline, which is a clear performance improvement.

On the other hand, there is a considerable performance gap between the CNN-Transformer and the baseline on the multi-pitch estimation and offset detection tasks. We argue that this is because both the ground-truth labels of multi-pitch estimation and offset detection have strong short-term temporal dependencies, and the current Transformer structure cannot easily model these relationships. For example, in the multi-pitch estimation task, it is quite common to observe a consequent sequence of note activations of the same pitch because of the duration of notes. Hence, if many neighboring frames of a particular frame have activation on a specific note pitch, a model should have prior knowledge to believe that activation is more likely to show up on this frame. As for the offset detection task, the offset location is also closely related to the energy decay over time of nearby preceding frames. In both cases, the model should give more weights to frames nearby when making decisions about a specific frame, i.e., short-term memory is crucial for the two tasks. Because of the forget gate inside the GRU unit, GRU tends to focus more on neighboring frames when training and inference. However, to enhance the ability to learn long-range dependencies, the Transformer is designed to connect all pairs of input and output positions, hence equally treating all time steps when doing detections. Therefore, the GRU can model this relationship more naturally than the Transformer. 

Table \ref{tab:res-subtasks} also includes the result of an ablation study to remove the positional encoding of the CNN-Transformers. It is worth noting that the positional encoding is crucial in all the subtasks, as the performance considerably degrades when it is removed. Based on this observation, we can infer that the absolute temporal location of each frame in the spectrogram is essential for the subsequent Transformer. 


Because the Transformer has considerable improvement on the velocity estimation performance, we incorporate the Transformer structure for the velocity branch in the following experiment.

\subsubsection{Final system}

Tables \ref{tab:res-frame} and \ref{tab:res-note} show the performance comparison of the proposed system and other state-of-the-art systems. We refer to the system in \cite{Kong2020amt} as HPT, the system with velocity Transformer, described in Section \ref{sec:method}, as HPT-T. In Table \ref{tab:res-frame}, As we expected, HPT-T outperforms the baseline HPT on the velocity task by 10\% lower mean absolute error and 6.8\% lower standard deviation. In the current system structure, the onset branch takes the transcription result produced by the velocity branch as a condition when training and inferring, so the performance of the velocity branch will impact the performance of the onset branch. On the onset detection task, although HPT-T is a little behind on the strict F1 measure ($-$0.35\%), it slightly outperforms the HPT on the F1 with tolerance by 0.08\%, which is a more important metric in note-level evaluation. We can infer that the information provided for the onset branch by the CNN-Transformer velocity branch is not as temporal-accurate as that of CNN-GRU, but more effective for the note-level onset detection. Also, the better onset branch benefits its downstream subtask, leading to better multi-pitch estimation performance. On the multi-pitch estimation task, HPT-T outperforms the HPT by 0.46\%. For the offset task, the improvement results from a better training strategy. As the result of the enhanced performance of each branch, the proposed system achieves better note-level results on all the note-level metrics, as shown in Table \ref{tab:res-note}. Although on the Note-3 and Note-4 metrics, HPT-T is slightly weaker than the model in \cite{Hawthorne2021} based a generic Transformer, our model has better performance on the Note-2 measure and provide satisfactory frame-level transcriptions as well, which is also a competitive result. 

Table \ref{tab:res-cross} shows the results of a cross-dataset evaluation on a subset of the MAPS database. Our system has a better frame-level F1 score than HPT and also performs better on Note-3 and Note-4 metrics. This indicates the proposed system has a reasonable ability to generalize to a dataset with different recording environments, and that the introduction of the Transformer does have positive effects on the system’s overall performance. 

\section{Conclusion}
\label{sec_conclusion}

We have explored the Transformer's ability to solve different AMT subtasks.
Based on our experiments, the proposed HPT-T system improves the transcription performance of the baseline on both frame-level and note-level metrics. We have further shown the decent generalization ability of our system by a cross-dataset evaluation. For future study, we plan to test more possible position representations in the current Transformer structure, instead of solely using the sinusoidal function as the positional encoding. In addition, we will try to restrict self-attention to prioritize neighboring frames of the respective output position to enhance the performance of the Transformer on multi-pitch estimation and offset detection tasks.

\vfill\pagebreak

\bibliographystyle{IEEEbib}
\bibliography{refs}

\begin{thebibliography}{10}

\bibitem{Benetos2019}
Emmanouil Benetos, Simon Dixon, Zhiyao Duan, and Sebastian Ewert,
\newblock ``Automatic music transcription: An overview,''
\newblock {\em IEEE Signal Processing Magazine}, vol. 36, no. 1, pp. 20--30,
  2019.

\bibitem{Vaswani2017}
Ashish Vaswani, Noam Shazeer, Niki Parmar, Jakob Uszkoreit, Llion Jones,
  Aidan~N. Gomez, {\L}ukasz Kaiser, and Illia Polosukhin,
\newblock ``Attention is all you need,''
\newblock in {\em Advances in Neural Information Processing Systems (NIPS)},
  2017, pp. 5998--6008.

\bibitem{Huang2019}
Cheng-Zhi~Anna Huang, Ashish Vaswani, Jakob Uszkoreit, Noam Shazeer, Ian Simon,
  Curtis Hawthorne, Andrew~M. Dai, Matthew~D. Hoffman, Monica Dinculescu, and
  Douglas Eck,
\newblock ``Music transformer: Generating music with long-term structure,''
\newblock in {\em International Conference on Learning Representations (ICLR)},
  2019.

\bibitem{Kong2020amt}
Qiuqiang Kong, Bochen Li, Xuchen Song, Yuan Wan, and Yuxuan Wang,
\newblock ``{High-resolution piano transcription with pedals by regressing
  onset and offset times},''
\newblock {\em IEEE/ACM Transactions on Audio Speech and Language Processing},
  vol. 29, pp. 3707--3717, 2021.

\bibitem{Hawthorne2019}
Curtis Hawthorne, Andriy Stasyuk, Adam Roberts, Ian Simon, Cheng Zhi~Anna
  Huang, Sander Dieleman, Erich Elsen, Jesse Engel, and Douglas Eck,
\newblock ``{Enabling factorized piano music modeling and generation with the
  Maestro dataset},''
\newblock in {\em International Conference on Learning Representations (ICLR)},
  2019, pp. 1--12.

\bibitem{Emiya2010}
Valentin Emiya, Nancy Bertin, Bertrand David, and Roland Badeau,
\newblock ``{MAPS - A piano database for multipitch estimation and automatic
  transcription of music},''
\newblock Research report, July 2010.

\bibitem{Marolt2004}
Matija Marolt,
\newblock ``A connectionist approach to automatic transcription of polyphonic
  piano music,''
\newblock {\em IEEE Transactions on Multimedia}, vol. 6, no. 3, pp. 439--449,
  2004.

\bibitem{Bock2012}
Sebastian B{\"{o}}ck and Markus Schedl,
\newblock ``{Polyphonic piano note transcription with recurrent neural
  networks},''
\newblock in {\em IEEE International Conference on Acoustics, Speech and Signal
  Processing (ICASSP)}, 2012, pp. 121--124.

\bibitem{Sigtia2016}
Siddharth Sigtia, Emmanouil Benetos, and Simon Dixon,
\newblock ``An end-to-end neural network for polyphonic piano music
  transcription,''
\newblock {\em IEEE/ACM Transactions on Audio, Speech, and Language
  Processing}, vol. 24, no. 5, pp. 927--939, 2016.

\bibitem{Hawthorne2018}
Curtis Hawthorne, Erich Elsen, Jialin Song, Adam Roberts, Ian Simon, Colin
  Raffel, Jesse Engel, Sageev Oore, and Douglas Eck,
\newblock ``{Onsets and frames: Dual-objective piano transcription},''
\newblock in {\em International Society for Music Information Retrieval
  Conference (ISMIR)}, 2018, pp. 50--57.

\bibitem{Kelz2019}
Rainer Kelz, Sebastian B{\"{o}}ck, and Gerhard Widmer,
\newblock ``{Deep polyphonic ADSR piano note transcription},''
\newblock in {\em IEEE International Conference on Acoustics, Speech, and
  Signal Processing (ICASSP)}, 2019, vol. 2019-May, pp. 246--250.

\bibitem{Kim2019}
Jong~Wook Kim and Juan~Pablo Bello,
\newblock ``{Adversarial learning for improved onsets and frames music
  transcription},''
\newblock in {\em International Society for Music Information Retrieval
  Conference (ISMIR)}, 2019, pp. 670--677.

\bibitem{Hawthorne2021}
Curtis Hawthorne, Ian Simon, Rigel Swavely, Ethan Manilow, and Jesse Engel,
\newblock ``{Sequence-to-sequence piano transcription with Transformers},''
\newblock in {\em International Society for Music Information Retrieval
  Conference (ISMIR)}, 2021.

\bibitem{Szeto2005}
Wai~Man Szeto, Kin~Hong Wong, and Chi~Hang Wong,
\newblock ``{Finding intensities of individual notes in piano music},''
\newblock in {\em Computer Music Modeling and Retrieval}, 2005.

\bibitem{Ewert2011}
Sebastian Ewert and Meinard M{\"{u}}ller,
\newblock ``{Estimating note intensities in music recordings},''
\newblock in {\em IEEE International Conference on Acoustics, Speech, and
  Signal Processing (ICASSP)}, 2011, pp. 385--388.

\bibitem{VanHerwaarden2014}
Sam {Van Herwaarden}, Maarten Grachten, and W.~{Bas de Haas},
\newblock ``{Predicting expressive dynamics in piano performances using neural
  networks},''
\newblock in {\em International Society for Music Information Retrieval
  Conference (ISMIR)}, 2014, pp. 47--52.

\bibitem{Jeong2017}
Dasaem Jeong and Juhan Nam,
\newblock ``{Note intensity estimation of piano recordings by score-informed
  NMF},''
\newblock in {\em AES International Conference}, 2017, pp. 124--131.

\bibitem{Jeong2018}
Dasaem Jeong, Taegyun Kwon, and Juhan Nam,
\newblock ``{A timbre-based approach to estimate key velocity from polyphonic
  piano recordings},''
\newblock in {\em International Society for Music Information Retrieval
  Conference (ISMIR)}, 2018, pp. 120--127.

\bibitem{Raffel2014}
Colin Raffel, Brian McFee, Eric~J. Humphrey, Justin Salamon, Oriol Nieto, Dawen
  Liang, and Daniel P.~W. Ellis,
\newblock ``{mir{\_}eval: A transparent implementation of common MIR
  metrics},''
\newblock in {\em International Society for Music Information Retrieval
  Conference (ISMIR)}, 2014, pp. 367--372.

\bibitem{Loshchilov2019}
Ilya Loshchilov and Frank Hutter,
\newblock ``{Decoupled weight decay regularization},''
\newblock in {\em International Conference on Learning Representations (ICLR)},
  2019.

\bibitem{Smith2017}
Leslie~N. Smith,
\newblock ``{Cyclical learning rates for training neural networks},''
\newblock in {\em IEEE Winter Conference on Applications of Computer Vision
  (WACV)}, 2017, pp. 464--472.

\end{thebibliography}

\end{document}